\documentclass[12pt]{article}
\usepackage{graphicx}
\def\pt{p_T}
\def\dis{distribution}

\begin{document} 
\begin{flushright}
OITS-797\\
June 2008
\end{flushright}
\vskip1cm
\begin{center}  {\Large {\bf Forward Production of Protons and Pions  \\
\vspace{.5cm}
 in Heavy-ion Collisions}}
\vskip .75cm
 {\bf Rudolph C. Hwa$^1$ and Li-Lin Zhu$^{1,2}$}
\vskip.5cm
{$^1$Institute of Theoretical Science and Department of
Physics\\ University of Oregon, Eugene, OR 97403-5203, USA\\
\bigskip
$^2$Institute of Particle Physics, Hua-Zhong Normal
University, Wuhan 430079, P.\ R.\ China}
\end{center}
\vskip.5cm

\begin{abstract} 
The problem of forward production of hadrons in heavy-ion collision at RHIC is revisited with modification of the theoretical treatment on the one hand and with the use of new data on the other. The basic formalism for hadronization remains the same as before, namely, recombination, but the details of momentum degradation and quark regeneration are improved. Recent data on the $p/\pi$ and $\bar p/p$ ratios are used to constrain the value of the degradation parameter. The $\pt$ spectrum of the average charged particles is well reproduced. A prediction on the $\pt$ dependence of the $\bar p/p$ ratio at $\eta=3.2$ is made.

\vskip0.5cm
PACS number:   25.75.Dw 

Keywords:  recombination, degradation, regeneration
\end{abstract}

\section{Introduction}

Theoretical study of hadron production in the forward direction in heavy-ion collisions is a difficult problem for several reasons. The empirical fact that the proton-to-pion ratio is large at large rapidity implies that neither fragmentation nor hydrodynamics can be successful in describing the process of hadron production in that region. Recombination is the natural hadronization mechanism for large baryon/meson ratio, but the parton momentum \dis\ at low $Q^2$ and large momentum fraction $x$  (contributing to hadronic Feynman $x_F$ in the range $0.3<x_F<1.0$) in nuclear collision is hard to determine, especially when momentum degradation and soft-parton regeneration cannot be ignored. The use of data as input to constrain unknown parameters is unavoidable; however, that is also where further complexity arises. Data on forward production at $\eta=3.2 \pm 0.2$ depend on both $p_L$ and $\pt$, resulting in a smearing of the $x$ \dis s of the partons that makes phenomenology difficult due to the inter-connectedness of all aspects of the dynamical problem. The problem was first studied in the framework of the recombination model in \cite{dh,rch,hy3} with the effects of parton regeneration taken into account  \cite{hy,hy2}.    The original data from PHOBOS show the \dis\ of charged particles at large $\eta$, but without $p_T$ measurement the value of $x_F$ cannot be determined \cite{phob}.  BRAHMS has measured both $\eta$ and $p_T$ dependences of charged hadrons \cite{ia}, but without particle identification the $p/\pi$ ratio cannot be inferred.  Very recently, there are preliminary data that indicate the $p/\pi$ ratio at $\eta = 3.2$ to be very large, $\sim 4$ at $p_T = 1.1$ GeV/c, 0-10\% centrality in Au-Au collisions at $\sqrt s=62.4$ GeV \cite{ps}, about 3 times higher than the prediction in \cite{hy2}.  The aim of this paper is to reexamine the problem of forward production and show that with appropriate changes in the treatment of degradation, regeneration and transverse momentum, there can be an understanding for the large $p/\pi$ ratio in the fragmentation region.

In addition to the new data on $p/\pi$ ratio there is also a new presentation of the $\bar{p}/p$ ratio by BRAHMS  for $\sqrt{s} = 62.4$ GeV, where the value of  $R_{\bar{p}/p} \simeq 0.02$ is given \cite{ia2}.  That value differs from the value $0.05$ inferred from the figure presented in \cite{hyc}, which was the value used in \cite{hy2}.  The new values of $R_{p/\pi}$ and $R_{\bar{p}/p}$ are consistent with the implication that there are more quarks or less antiquarks than what were obtained in \cite{hy2}.  That provides a hint for us to look for the area in the formalism where the treatment of degradation and regeneration may be improved.  Regeneration is an effect that depends on momentum degradation in forward propagation, which in turn depends on the degradation parameter $\kappa$ that is not known except by fitting the data.  With new data available, the whole procedure needs to be revised. In this paper we change the strategy of our phenomenology in order to take advantage of   the additional constraints provided by the particle ratios.

The formalism for forward production is basically the same as discussed in \cite{hy,hy2}.  We describe its essence in Sec.\ 2, but with special emphasis on changes that are necessary to improve the treatment. In Sec.\ 3 momentum degradation and  quark regeneration are investigated with significant changes from \cite{hy,hy2}. How the data on particle ratio can be used to constrain $\kappa$ is discussed in Sec.\ 4, followed by consideration of the transverse momentum in Sec.\ 5. Conclusion is given in Sec.\ 6.

\section{Basic Formalism for Forward Production}

In the recombination model (RM) \cite{dh,rch,hy3} hadron production can be described by the basic equations
\begin{eqnarray}
H^{AB}_p (x) &=& \int {dx_1\over x_1} {dx_2\over x_2} {dx_3\over x_3}F^{AB}_{uud}(x_1, x_2, x_3)R_p(x_1, x_2, x_3, x)\ ,   \label{1}\\
H^{AB}_{\pi}(x) &=& \int {dx_1  \over 
x_1}{dx_2 \over  x_2} F^{AB}_{q\bar{q}} (x_1, x_2) R_{\pi}(x_1, x_2, x) \ .   \label{2}
\end{eqnarray}
for proton and pion, respectively, where only one-dimensional consideration is needed for forward production, with $x\equiv x_F = 2p_L/\sqrt{s}$ for hadron, and $x_i$ being the  momentum fractions of partons \cite{hy,hy2}.  The recombination functions (RF), $R_p$ and $R_{\pi}$, depend on the wave functions of the hadrons and are summarized in \cite{hy}.  The major task to render Eqs.\ (\ref{1}) and (\ref{2}) useful is to determine the parton distributions $F^{AB}_{uud}$ and $F^{AB}_{q\bar{q}}$ for the problem at hand.  For forward production the largest contribution can be attained if the quarks arise from different initial nucleons so that their momenta do not have to be shared among the quarks originating from the same nucleon.  That means $F^{AB}_{uud}(x_1, x_2, x_3)$ depends on a factorizable product of independent quark distributions $F_q^{\nu_i}(x_i)$ at momentum fraction $x_i$ of an incident nucleon after $\nu_i$ collisions with the target nucleus $B$; similarly, $F^{AB}_{q\bar{q}}(x_1,x_2)$ involves quark and antiquark distributions.  If only 3 or 2 nucleons in the projectile $A$ are considered in each collision, we can define the $p$ and $\pi$ distributions from such sources as $H^{(3)B}_p$ and $H^{(2)B}_{\pi}$, which are then related to the overall distributions for $AB$ collision by 
\begin{eqnarray}
H^{AB}_p (x,b) = \int {d^2s  \over  \sigma}
{\left[\sigma T_A(s)  \right]^3\over 3!}\,
H^{(3)B}_p (x, b, s) \ .
\label{3}
\end{eqnarray}
\begin{eqnarray}
H^{AB}_{\pi} (x,b) = \int {d^2s  \over  \sigma}
{\left[\sigma T_A(s)  \right]^2\over 2!}  H^{(2)B}_{\pi} (x, b, s) \ ,
\label{4}
\end{eqnarray}
where $b$ is the impact parameter. These formulas are derived in \cite{hy}.  Clearly to describe $H^{(3)B}_p$ and $H^{(2)B}_{\pi}$ is a simpler problem than in Eqs.\ (\ref{1}) and (\ref{2}), since the corresponding parton distributions are for 3 and 2 nucleons, respectively, in the projectile.  In Eqs.\ (\ref{3}) and (\ref{4}) $\sigma$ denotes the inelastic cross section of nucleon-nucleon collision, and $T_A(s)$ is the thickness function for a tube in $A$ at impact parameter $s$.  

The recombination equation for the reduced projectile going through the target $B$ is as in  Eqs.\ (\ref{1}) and (\ref{2})
\begin{eqnarray}
H^{(3)B}_{p} (x,b,s) = \int {dx_1 \over x_1} {dx_2 \over x_2}{dx_3 \over x_3} F^{(3)B}_{qqq} (x_1, x_2, x_3; |\vec{s}- \vec{b}|)R_p(x_1, x_2, x_3,x)
\label{5}
\end{eqnarray}
\begin{eqnarray}
H^{(2)B}_{\pi} (x,b,s) = \int {dx_1 \over x_1} {dx_2 \over x_2} F^{(2)B}_{q\bar{q}} (x_1, x_2;  |\vec{s}- \vec{b}|)R_{\pi}(x_1, x_2, x) \quad ,
\label{6}
\end{eqnarray}
where $F^{(3)B}_{qqq} (x_1, x_2, x_3; |\vec{s}- \vec{b}|)$ is the 3-quark joint distribution after 3 nucleons transverse  the target nucleus at impact parameter $|\vec{s}- \vec{b}|$ in $B$.  We shall neglect the minor flavor dependence of nucleons and quarks in the following.  Similarly, $ F^{(2)B}_{q\bar{q}} (x_1, x_2;  |\vec{s}- \vec{b}|)$ is the $q\bar{q}$ distribution after 2 nucleons go through $B$.       In Ref.\ \cite{hy}    $F^{(3)B}_{qqq}$  is assumed to have a factorizable form.  We now give a derivation of that form and in the process determine the appropriate average number of collisions.  The same follows for $F^{(2)B}_{q\bar{q}}$.

If each nucleon in the projectile nucleus $A$ at $\vec{s}$ makes on average $\bar{\nu}$ collisions in $B$, where 
\begin{eqnarray}
\bar{\nu} \equiv  \bar{\nu}^{pB}(|\vec{s}-\vec{b}|) =  {\sigma T_B (|\vec{s}-\vec{b}|) \over 1 - \exp \left[ - \sigma T_B (|\vec{s}-\vec{b}|)\right]} \ ,
\label{7}
\end{eqnarray}
then 3 nucleons make on average $3\bar{\nu}$ collisions.  Assuming a Poisson distribution in $\nu$, we have
\begin{eqnarray}
H^{(3)B}_p (x; b, s) = \sum_{\nu} H^{(3)B}_p (x; \nu) P_{3\bar{\nu}}(\nu) \quad ,
\label{8}
\end{eqnarray}
where
\begin{eqnarray}
P_{\bar{\nu}}(\nu) = {\bar{\nu}^{\nu}  \over  \nu!} e^{-\bar{\nu}} \ .
\label{9}
\end{eqnarray}
Applying Eq.\ (\ref{8}) to (\ref{5}) the sum over $\nu$ can be moved past the integrals and we can write
\begin{eqnarray}
F^{(3)B}_{qqq} (x_1, x_2, x_3; |\vec{s}- \vec{b}|) = \sum_{\nu} F^{(3)B}_{qqq} (x_1, x_2, x_3; \nu) P_{3\bar{\nu}} (\nu)\ .
\label{10}
\end{eqnarray}
Now, $\nu$ is the total number of wounded nucleons experienced by the target nucleus $B$, irrespective of how it is distributed among the incident nucleons.  With three such nucleons that are independent, we have
\begin{eqnarray}
F^{(3)B}_{qqq} (x_1, x_2, x_3; \nu) = {1 \over 3^{\nu}}\sum_{\nu_1, \nu_2, \nu_3}{\nu! \over \nu_1! \nu_2! \nu_3!} F^{\nu_1}_{q} (x_1) F^{\nu_2}_{q}(x_2) F^{\nu_3}_{q} (x_3) \ ,
\label{11}
\end{eqnarray}
where the summation over $\nu_i$ is constrained by $\sum_i \nu_i = \nu$, each starting from $\nu_i = 0$.  Each term in the summand is a product of single-quark distributions in a proton that has undergone $\nu_i$ collisions with the target.  They include the effects of degradation and regeneration to be discussed below.

The use of $3\bar{\nu}$ in the Poisson distribution in Eq.\ (\ref{10}) is based on the assumption that all three nucleons in the projectile are lined up in the same tube at impact parameter $|\vec{s}-\vec{b}|$ in $B$, since otherwise the forward partons are not nearby in the transverse plane and are unlikely to recombine to form a proton.  Thus the same $\bar{\nu}^{pB}$ applies to each of the three nucleons.  Substituting Eq.\ (\ref{11}) into (\ref{10}) and making use of the implicit  $\delta _{\nu, \nu_1 + \nu_2 +\nu_3}$ contained in the summation in (\ref{11}), the sum over $\nu$ can readily be carried out, yielding 
\begin{eqnarray}
F^{(3)B}_{qqq} (x_1, x_2, x_3; |\vec{s}- \vec{b}|) = \prod^3_{i = 1} F^{\bar\nu}_q (x_i) \ ,
\label{12}
\end{eqnarray}
where 
\begin{eqnarray}
 F^{\bar\nu}_q (x_i)=  \sum^{\infty}_{\nu_i = 0} F^{\nu_i}_q (x_i) P_{\bar{\nu}}(\nu_i)
\label{13}
\end{eqnarray}
with $\bar{\nu}$ being defined in Eq.\ (\ref{7}) for a $pB$ collision.  Using Eq.\ (\ref{12}) in (\ref{5}) and then in (\ref{3}) we have reduced the proton production problem in $AB$ collision to the only issue at hand, i.e., how the parton distribution $F^{\bar{\nu}}_q (x_i)$ is to be determined.

For forward production we ignore the production of resonances and their decays. Proton is in the symmetric state in $SU(2)\times SU(2)$ for (spin, isospin). In $2\times 2\times 2=4+2_a+2_s$ for $qqq$, the symmetric state $(2_s,2_s)+(2_a,2_a)$ is 8 out of a total of 64 states, so the statistical factor $g_{st}$ in $R_p(x_1,x_2,x_3,x)$ is 1/8. For pion there is no change in $R_{\pi}(x_1,x_2,x)$ from that given in \cite{hy}.

\section{Momentum Degradation and Quark Regeneration}

The problem of forward production in $pB$ collision has been treated in the framework of the valon model, which connects the bound-state problem of a static proton (in terms of constituent quarks) with the structure problem of a proton in collision (in terms of partons) \cite{rch,hy3,hy4}.  Without momentum degradation the quark distribution in a free proton is given by
\begin{eqnarray}
 F_q (x_i,Q^2)=  \int^{1}_{x_i} dy G(y) K\left({x_i\over y} , Q^2 \right) \ ,
\label{14}
\end{eqnarray}
where $G(y)$ is the valon distribution, $y$ being the momentum fraction of the valon, and $K(z, Q^2)$ is the quark distribution in a valon, both of which have been parameterized and updated in \cite{hy5}.  With momentum degradation in proton-nucleus collision both $G(y)$ and $K(z, Q^2)$ are modified, as described in \cite{hy,hy2}.  However, we have come to the realization that Eq.\ (\ref{14}) itself needs modification, a new development which we now describe from the beginning.

A proton has three valons, which are the constituent quarks in the bound-state problem.  When a proton wounds $\nu$ nucleons in the target nucleus, it does not matter which of the the 3 valons causes the wounding; they can act independently.  It is important to recognize the possibility that one of the valons may not undergo any momentum degradation, while the other two are responsible for causing $\nu$ wounded nucleons in the target.  Although the probability of that is low, the valence quark in the undegraded valon would have higher momentum.  The point is that we should consider all possibilities, which can be expressed in the form
\begin{eqnarray}
 F^{\nu}_q (x_i)= {1 \over 2^{\nu}} \sum^{\nu}_{\mu=0} {\nu ! \over \mu! (\nu - \mu)!}  \int^{\kappa^{\mu}}_{x_i} dy' G^{'}_{\mu}(y') K\left({x_i \over y'}\right) \ ,
 \label{15}
\end{eqnarray}
where the $Q^2$ dependence, shown explicitly in Eq.\ (\ref{14}), is suppressed because it is at some unspecified low value that is not of central importance here.  $G^{'}_{\mu}(y')$ is the modified valon distribution due to degradation to be discussed below, together with the upper limit of integration.  The Poissonian averaging of $\mu$, the number of nucleons in $B$ wounded by a valon, allows $\mu$ to be zero, while the total number of wounded nucleons is fixed at $\nu$.  Thus the way that the valons are treated in a projectile nucleon is analogous to the way that the nucleons are treated in a projectile nucleus.

If the momentum fraction that a valon retains after a collision with a nucleon in $B$ is $\kappa$, then after $\mu$ collisions the modified valon distribution is 
\begin{eqnarray}
y'G'_{\mu}(y') = \int^1_{y'}  dy G(y)\kappa ^{\mu} \delta \left({ y'  \over y} - \kappa ^{\mu} \right) \  ,
 \label{16}
\end{eqnarray}
which satisfies the normalization condition
\begin{eqnarray}
\int dy'G'_{\mu}(y') = \int dy G(y) = 1 \  .
\label{17}
\end{eqnarray}
The solution of Eq.\ (\ref{16}) is 
\begin{eqnarray}
G'_{\mu}(y') = \kappa ^{- \mu}\ G\left( \kappa ^{-\mu} y' \right) \  .
\label{18}
\end{eqnarray}
It is clear that the maximum value of $y'$ is $\kappa^{\mu}$ because of the $\mu$-fold degradation, thus setting the upper limit of integration in Eq.\ (\ref{15}).  Furthermore, the average momentum of the degraded valon is
\begin{eqnarray}
\left<y' \right>_{\mu} =  \int dy' y' G'_{\mu}(y')  =   \kappa ^{\mu}   \left<y\right> = {1 \over 3}  \kappa ^{\mu}      \  ,
\label{19}
\end{eqnarray}
where $ \left<y\right>$ is the average momentum fraction of a valon in a free proton and is $1/3$.  Thus Eq.\ (\ref{19}) expresses the effect of degradation in this simple model of multiplicative momentum loss of the sequential collision process.

The valence quark distribution in a proton after $\nu$ collision is as expressed in Eq.\ (\ref{15}), but with $K(z)$ replaced by the non-singlet component $K_{NS}(z)$, which is specified in \cite{hy5}.  Due to the $\mu$ dependence of $G'_{\mu}(y')$ in Eqs.\ (\ref{18}) and (\ref{19}), the sum over $\mu$ in (\ref{15}) acquires special significance at low $\mu$, as remarked earlier before that equation.  It is the $\mu = 0$ term that renders the valence quark distribution at intermediate $x_i$ insensitive to the value of $\kappa$.  In this respect our treatment here is an improvement over that in \cite{hy,hy2}.

For the regenerated sea quark distributions the earlier treatment can also be improved.  In \cite{hy2} the quark distribution $K(z)$ in a valon is written in the two-component form
\begin{eqnarray}
 K(z) = K_{NS}(z) + L'' (z) \ ,
\label{20}
\end{eqnarray}
where $L'' (z)$ represents the regenerated sea quark distribution in a valon, including gluon conversion.  
The regenerated $\bar q$ \dis, $F_{\bar q}^\nu(x_i)$, for a nucleon making $\nu$ collisions with the target is then as given in Eq.\ (\ref{15}), but with $K(z)$ replaced by $L''(z)$. We now realize that such a convolution equation gives only a part of the total $\bar q$ \dis\ because the momentum lost by a nucleon after $\nu$ collisions is not totally accounted for by that convolution equation.
  The average momentum loss of a nucleon as a fraction of the initial momentum after $\nu$ collisions is $1 - \left<x \right>_{\nu}$, where
\begin{eqnarray}
\left<x \right>_{\nu} = {1 \over 3^{\nu}} \sum_{\mu _1, \mu _2, \mu_3}  {\nu ! \over \mu _1! \mu _2! \mu_3!} \kappa^{\mu_1}\kappa^{\mu_2}\kappa^{\mu_3} = \kappa^\nu  \ .
\label{21}
\end{eqnarray}
We assume that the momentum loss is converted totally to $u\bar{u} + d\bar{d}$ pairs without strange quarks.  Thus the regenerated $\bar{q} \, (\bar{u}$ or $\bar{d})$ distribution for each nucleon in the projectile, $F^{\nu}_{\bar{q}}(x)$,  should satisfy the sum rule (with the subscript $i$ on $x_i$ suppressed)
\begin{eqnarray}
\int dx F^{\nu}_{\bar{q}}(x) = {1 \over 4}\left( 1 - \kappa^{\nu} \right) \ .
\label{22}
\end{eqnarray}
We adopt the approximate form for the $x$ dependence 
\begin{eqnarray}
F^{\nu}_{\bar{q}}(x) =  f_{\nu} (1 - x)^n    \ ,
\label{23}
\end{eqnarray}
so $f_{\nu} = (1 - \kappa^{\nu})/4(n +1) $.  We shall use $n = 7$, since that is suggested by the $\bar{q}$ parton distribution of a free nucleon for $x$ not too small and at low $Q^2$.  For the values of $\kappa$ and $\nu$ that we encounter below, Eq.\ (\ref{23}) gives values of $F^{\nu}_{\bar{q}}(x)$, for $x > 0.2$, far greater than those obtained by the convolution of $G'_{\nu}(y')$ with $L''(x/y')$, as determined in \cite{hy2}; the latter is therefore neglected hereafter. 

To summarize, for quark ($u$ or $d$) distribution in $pB$ collision after $\nu$ wounded nucleons in $B$, we have
\begin{eqnarray}
F^{\nu}_{q}(x) =  F^{\nu}_{q_v}(x) + F^{\nu}_{\bar{q}}(x)    \ ,
\label{24}
\end{eqnarray}
where $F^{\nu}_{q_v}(x)$ is the valence quark distribution given by Eq.\ (\ref{15}) with $K_{NS}(z)$ in place of $K(z)$, and $F^{\nu}_{\bar{q}}(x)$ is the regenerated quark distribution given by Eq.\ (\ref{23}).  The antiquark distribution is, of course, just the second term in (\ref{24}).

\section{Particle Ratios}

Having obtained the modified quark distribution due to degradation and regeneration, we can now use Eq.\ (\ref{24}) in (\ref{13}) for the $i$th nucleon, and then in (\ref{12}) for $qqq$ distribution emerging from 3-nucleons colliding with target $B$ at impact parameter $|\vec{s}- \vec{b}|$.  That result can then be used in Eqs.\ (\ref{3}) and (\ref{5}) to determine the $x$ distribution of produced proton in $AB$ collision.  Exactly the same procedure can be followed to obtain the spectra of $\pi$ and $\bar{p}$ with appropriate use of the the $\bar{q}$ distribution for $q\bar{q}$ and $\bar{q}\bar{q}\bar{q}$ recombination.

We show in Fig.\ 1 the results of our calculation of  the $x$ distributions of $p$, $\pi$, and $\bar{p}$ for $\kappa = 0.7$, and $b=3.3$ fm for 0-10\% centrality in Au-Au collisions. The value of $\kappa$ is chosen for reasons to be given below.  Evidently, the $p$ distribution is much higher than the other two for $x>0.5$, since it is due to the recombination of three valence quarks from three different nucleons in the projectile $A$.  Moreover, it decreases more slowly with increasing $x$ due to the slower decrease of valence quark distribution compared to the sea quarks.  Thus the $p/\pi$ ratio is large and increases with increasing $x$. The $\pi$ distribution is much higher than the $\bar{p}$ distribution, because of the effect of valence quark in $\pi$ that is lacking in $\bar{p}$.  Similar plots can also be made for other values of $\kappa$, but in the absence of any data on the hadronic $x$ distributions 
the comparison among different $\kappa$ values can better be presented in a different format, as shown below. The general trend is that lower $\kappa$ leads to higher level of $\bar q$ and therefore higher $\pi$ and $\bar p$ at low $x$.

Recently, data have become available on the particle ratios of both $\bar{p}/p$ \cite{ia2} and $p/\pi$
\cite{ps}.  It is then very revealing for us to make parameteric plots of those ratios  for various values of $x$ and $\kappa$. We use Eqs.\ (\ref{1}) and (\ref{2}) to 
calculate $H_h^{AB}(x, \kappa)$ for $b=3.3$ fm and $h=p, \pi, \bar p$ and show their ratios $H_{\bar p}/H_p$ versus $H_p/H_\pi$ in Fig.\ 2, in which the grid lines are for constant $x$ (in solid lines) and constant $\kappa$ (in dashed lines).  
 It is clear that all lines have negative slopes in that figure because $\bar{q}$ is involved in the numerator of $H_{\bar p}/H_p$, but in the denominator of $H_{p}/H_\pi$.  Large values of $H_{p}/H_\pi$  can be achieved only when $x$ is $> 0.4$; that is the region where the valence quarks dominate and the sea quarks are suppressed.  At fixed $x$ both ratios depend sensitively on $\kappa$, more so for  $H_{\bar p}/H_p$ than for  $H_{p}/H_\pi$, because of the number of $\bar{q}$ involved.  The smaller $\kappa$ is, the more degradation there is, and the regenerated $\bar{q}$ boosts  $H_{\bar p}/H_p$ and suppresses $H_{p}/H_\pi$.

The data on $R_{\bar{p}/p}$ and $R_{p/\pi}$ depend on the values of $p_T$ at which the hadrons are included in the determination of the ratios.  
$R_{h'/h}$ cannot be identified with $H_{h'}/H_h$ until after the $p_T$ \dis\ is considered, a topic to be discussed in the next section.
  So far we have   only treated the dynamical processes that lead to the $x$ distributions. At fixed $\eta$ the longitudinal $p_L$ and the transverse $p_T$ are, of course, not kinematically independent.  The range of $x$ that is phenomenologically relevant to our study should correspond to the range of $p_T$ in which the experimental values of the particle ratios are determined.  Since the mismatch between $R_{h'/h}$ and $H_{h'}/H_h$ is not large, as we shall discuss later, let us here mark on Fig.\ 2 the data point that corresponds to \cite{ps,ia2}
\begin{eqnarray}
R_{p/\pi} = 4.08 \pm 0.2,\qquad \eta = 3.2 \pm 0.2,\qquad 0.9 < p_T < 1.3\ \rm{GeV/c}    \ ,
\label{25}
\end{eqnarray}
\begin{eqnarray}
R_{\bar{p}/p} = 0.0231 \pm 0.0012,\qquad y = 3.0 \pm 0.1,\qquad  0.5 < p_T<1.4\ \rm{GeV/c}   \ .
\label{26}
\end{eqnarray}
The grid lines in Fig.\ 2 then suggest that the relevant values of $x$ and $\kappa$ are
\begin{eqnarray}
x \simeq 0.55 \qquad  \rm{and} \qquad \kappa \simeq 0.67   \ ,
\label{27}
\end{eqnarray}
For that reason  the $x$ distributions in Fig.\ 1 are shown for $\kappa = 0.7$.

What we have done so far is essentially the first step of an iteration process, in which the focus is on the $x$ \dis. The next step is to consider the transverse momentum based on the result of the first step and to improve on the overall phenomenology.

\section{Transverse Momentum}

The $p_T$ dependence of the produced particles has been discussed in \cite{hy2}.  Let us first give a summary of that consideration.  Since hard scattering is suppressed in the fragmentation region, we ignore shower partons for $x > 0.2$.  This approximation is supported by the data on the $p_T$ distribution of charged particles at $\eta = 3.2$ \cite{ia}, which shows an exponential behavior for $p_T$ up to 2 GeV/c without up-bending due to power-law behavior.  Thus we write the $x$ and $p_T$ distributions of a produced hadron $h$ in the factorizable form
\begin{eqnarray}
{x \over p_T} {dN_h \over dxdp_{T}} =  H_h(x,\kappa) V_h(p_T) ,
\label{28}
\end{eqnarray}
where the transverse part is normalized by 
\begin{eqnarray}
\int_0^\infty dp_T\ p_T\ V_h(p_T) = 1 \ ,    
\label{29}
\end{eqnarray}
rendering
\begin{eqnarray}
x{dN_h\over dx}=H_h(x,\kappa)   \ ,
\label{30}
\end{eqnarray}
which is our starting point in Eqs.\ (\ref{1}) and (\ref{2}).  

The properties of $V_h(\pt)$ described in \cite{hy2} are adapted from the treatment of $\pt$ \dis\ in central collisions at mid-rapidity for which the only recombination process is in the transverse plane \cite{hy6}. Here, we have treated in detail the degradation, regeneration and recombination of partons in the forward production, so it is inappropriate to append a separate recombination of thermal partons with independent recombination functions for the transverse component. Since no shower partons are involved, we shall simply take a common exponential form for all hadrons, but allowing the inverse slopes $T_h$ to differ for hadrons with different masses, as suggested by hydrodynamical flow. Thus we write
\begin{eqnarray}
V_h(\pt)={1\over 2T_h^2} e^{-\pt/T_h}   \label{31}
\end{eqnarray}
with normalization chosen to satisfy Eq.\ (\ref{29}). We parametrize $T_h$ by
\begin{eqnarray}
T_h=T_0+m_h \langle v_t\rangle^2  \ ,    \label{32}
\end{eqnarray}
where the second term expresses the flow contribution. Since at large $x$ the dominant momentum direction is longitudinal, the mass-dependent component of the transverse momentum is expected to be small compared to the thermal component characterized by $T_0$.

Although $x$ and $p_T$ appear independent in Eq.\ (\ref{28}), they are kinematically constrained when $\eta$ is fixed.  They are related by
\begin{eqnarray}
x =  {2p_T\over \sqrt{s}} \sinh\eta \  .
 \label{33}
\end{eqnarray}
At $\eta = 3.2$, the range $1 \leq p_T \leq 1.5$ GeV/c corresponds to $0.39 \leq x \leq 0.59$.  On the other hand, if the rapidity $y$ is fixed,  the relationship depends on the particle mass.  At $y = 3.0$, the range $1 \leq p_T \leq 1.5$ GeV/c corresponds to $0.32 \leq x_{\pi}\leq 0.48$ and $0.44 \leq x_p \leq 0.57$.  The value $x \simeq 0.55$ determined from our theoretical grid lines in Fig.\ 2 lies within the range of $x$ values above for the data on $R_{p/\pi}$ at $\eta = 3.2$ in Eq.\ (\ref{25}) and also within the range of $x_p$ values above for the data on  $R_{\bar{p}/p}$ at $y = 3.0$ in (\ref{26}).  This is a non-trivial achievement, since the formalism described in Sec.\ 3 makes no reference to $p_T$, so the grid lines for
the ratios of $H_{h'/h}(x, \kappa)$ at constant
 $\kappa$ and $x$ need not imply any $\pt$ values that correspond to the relevant $x$ and $\pt$ values of the experimental $R_{h'/h}$ at fixed $\eta$.                       

The $\pt$ \dis\ given in \cite{ia} is to be identified with our calculation as follows
\begin{eqnarray}
{dN\over 2\pi\pt d\pt d\eta}={1\over 2\pi}H_{h}(x) V_{h}(\pt) \ ,   \label{34}
\end{eqnarray}
since upon integration over $\pt d\pt d\phi$ and using  Eq.\ (\ref{29}) it yields $H_{h}(x)$. Strictly speaking, holding $\eta$ fixed on the LHS is not the same as holding $x$ fixed on the RHS. But the data are analyzed at $\eta=3.2\pm 0.2$, so there are bands of $\eta$ and $x$ values in which Eq.\ (\ref{34}) is approximately valid. 
The data on $R_{p/\pi}(\pt)$ are then to be related to our calculation by
\begin{eqnarray}
R_{p/\pi}(\pt) = {H_p[x(\pt)]\over H_{\pi}[x(\pt)]}  {V_p(\pt)\over V_{\pi}(\pt)}  \ ,   \label{35}
\end{eqnarray}
where the ratio $H_p/H_{\pi}$ is to be determined by fixing $\eta=3.2$ and $\kappa=0.67$. In Fig.\ 3 we show that ratio by the dashed line, which has a significant $\pt$ dependence. Furthermore, the average magnitude of $H_p/H_{\pi}$ accounts for the major part of $R_{p/\pi}$, and cannot arise without a realistic treatment of degradation and regeneration. The reason for the dashed line to increase with $\pt$ is that at fixed $\eta$ higher $\pt$ means higher $x$, where $\bar q$ is suppressed compared to $q$, resulting in $\pi$ being suppressed relative to $p$. The solid line in Fig.\ 3 includes the effect of $V_p/V_\pi$, which we get from Eqs.\ (\ref{31}) and (\ref{32})
\begin{eqnarray}
{V_p(\pt)\over V_\pi(\pt)}=\left({T_\pi\over T_p}\right)^2 \exp \left[-\pt\left({1\over T_p}-{1\over T_\pi}\right)\right ] \ .   \label{36}
\end{eqnarray}
Since the $m_h\langle v_t\rangle^2$ term in Eq.\ (\ref{32}) is small compared to $T_0$, as shall show presently, the above ratio is approximately $\exp[(m_p-m_\pi)\pt\langle v_t\rangle^2/T_0^2]$, which shows the effect of mass difference in elevating the dashed line to the solid line. The result of fitting the data \cite{ps} gives
\begin{eqnarray}
\langle v_t\rangle^2/T_0^2 = 0.7\ ({\rm GeV/c})^{-2} \ .   \label{37}
\end{eqnarray}
This is consistent with $m_p\langle v_t\rangle^2\ll T_0$, when $T_0$ is 0.2 GeV/c to be determined below.

The $\pt$ \dis\ itself is an additional test of our model, since the absolute normalization is not canceled as in a ratio.  The data \cite{ia} are for all  charged hadrons without particle identification, for which we  treat $(h^+ + h^-)/2$ as $h^{\pm} = \left[p + \bar{p} + 1.2 \left(\pi^+ + \pi^-\right) \right]/2$, where the $K/\pi$ ratio of $\simeq 0.2$ is used \cite{ia2}. As the third step in our iteration process, we calculate $H_{h^\pm(x)}V_{h^\pm}(\pt)/2\pi$ holding $x$ and $\kappa$ fixed as in Eq.\ (\ref{27}) and adjust $T_0$ to fit the data according to Eq.\ (\ref{34}).  The result is shown in Fig.\ 4 for $T_0=200$ MeV; it agrees with the data very well. Since the normalization is fixed by the $H_h(x,\kappa)$ functions and is not adjustable,  a good fit is remarkable. 

Putting the obtained value of $T_0$ in Eq.\ (\ref{37}), we have
\begin{eqnarray}
T_0=0.2\ {\rm GeV/c},  \qquad  \langle v_t\rangle^2=0.028.  \label{38}
\end{eqnarray}
The value of $\langle v_t\rangle\simeq 0.17$ seems reasonable in view of the dominance of longitudinal expansion in the fragmentation region. The significance of this work is, of course, not in the transverse aspect of the problem, but on the longitudinal momentum distributions of the forward particles, which affect the $\pt$ \dis. The large $p/\pi$ ratio found in the BRAHMS data at $\eta=3.2$ cannot be understood without a proper treatment of the $x$ \dis s in the fragmentation region.

As a prediction of this work, we can calculate the $\pt$ dependence of the $\bar p/p$ ratio at fixed $\eta=3.2$. Since $V_{\bar p}/V_p=1$ for $h=\bar p, p$, only $H_h[x(p_T)]$ contributes to the ratio $R_{\bar p/p}(p_T)$. Using Eq.\ (\ref{33}), we have
\begin{eqnarray}
R_{\bar p/p}(\pt) = {H_{\bar p}[x(\pt)]\over H_{p}[x(\pt)]} \ .   \label{39}
\end{eqnarray}
The result for  ${\sqrt s}=62.4$ GeV, $b=3.3$ fm and $\kappa=0.67$ is shown in Fig.\ 5. The range of $\pt$ covered by the plot corresponds to $x$ roughly between 0.3 and 0.6 at $\eta=3.2$. Note that the result is for fixed $\eta$, not fixed $y$. The reason for the decrease of $R_{\bar p/p}(p_T)$ with increasing $\pt$ is clearly the increase of $x$ where $\bar q$ at momentum fraction $x_i$, approximately $x/3$, becomes more suppressed than $q$ at the same $x_i$.  
 A verification of this prediction would lend further support to our model.

\section{Conclusion}

This work differs from the earlier attempt in \cite{hy2} in three important ways. Firstly, new data are available that put more stringent constraints on unknown parameters. Secondly, significant modifications have been made in the treatment of degradation, regeneration, and transverse momenta. Thirdly, the order that phenomenology is carried out is reversed due to the new empirical knowledge about the particle ratios. Using $p/\pi$ and $\bar p/p$ ratios as input, we are able to determine the degradation parameter $\kappa$, which enables us to calculate the $x$ \dis s of the hadrons. At fixed $\eta$ that implies a $\pt$ dependence of the $p/\pi$ ratio arising from the $x$ \dis s; that $\pt$ dependence accounts for a large part of the data on that ratio, the balance being due to the exponential $\pt$ \dis s that are mass dependent. In fitting the particle ratio the calculated result is insensitive to the absolute normalization of the yield. The latter is shown to be correct when we succeed in reproducing the $\pt$ spectrum of the average charged particle. That is a significant achievement because the yields of protons, pions and antiprotons at large $\eta$ depend strongly on the dynamical process of momentum degradation and soft-parton regeneration.

Although the degradation parameter $\kappa$ is determined by data fitting, to get the spectra correctly for all hadrons through one such parameter relies on the validity of the treatment of the various subprocesses. Our results suggest that our model has captured the essence of the dynamics involved. In particular, the large $p/\pi$ ratio would not have emerged from our calculation if recombination has not been used as the mechanism for hadronization.

Since proton production at large $x$ is due to the recombination of three valence quarks from three nucleons in the projectile, in which there are numerous other valence quarks from other nucleons, we do not expect the events triggered by a large-$x$ proton would have correlated partners distinguishable from the background. In that respect the hadronization problem is similar to that at intermediate $\pt$ in heavy-ion collision at LHC, where so many semi-hard jets are produced that shower partons are dense and can recombine with large $p/\pi$ ratio \cite{hy7}. For the same reason as at large $x$ studied here, it was also predicted that for triggers in the $10<\pt<20$ GeV/c range no correlation structure of associated particles would be found. Thus to a certain extent what we can learn about forward production at RHIC may reveal some aspects of the characteristics of what may be observed at intermediate $\pt$ at midrapidity at LHC.

\section*{Acknowledgment}

We are grateful to I.\ C.\ Arsene, P.\ Staszel, F.\ Videbaek, and C.\ B.\ Yang for helpful communication. This work was supported,  in
part,  by the U.\ S.\ Department of Energy under Grant No. DE-FG02-96ER40972 and by the National Science Foundation in China under Grant 10775057
and by the Ministry of Education of China under Grant No. 306022 and project IRT0624.

\newpage
\begin{figure}[htbp]
\centering
\includegraphics[width=6in]{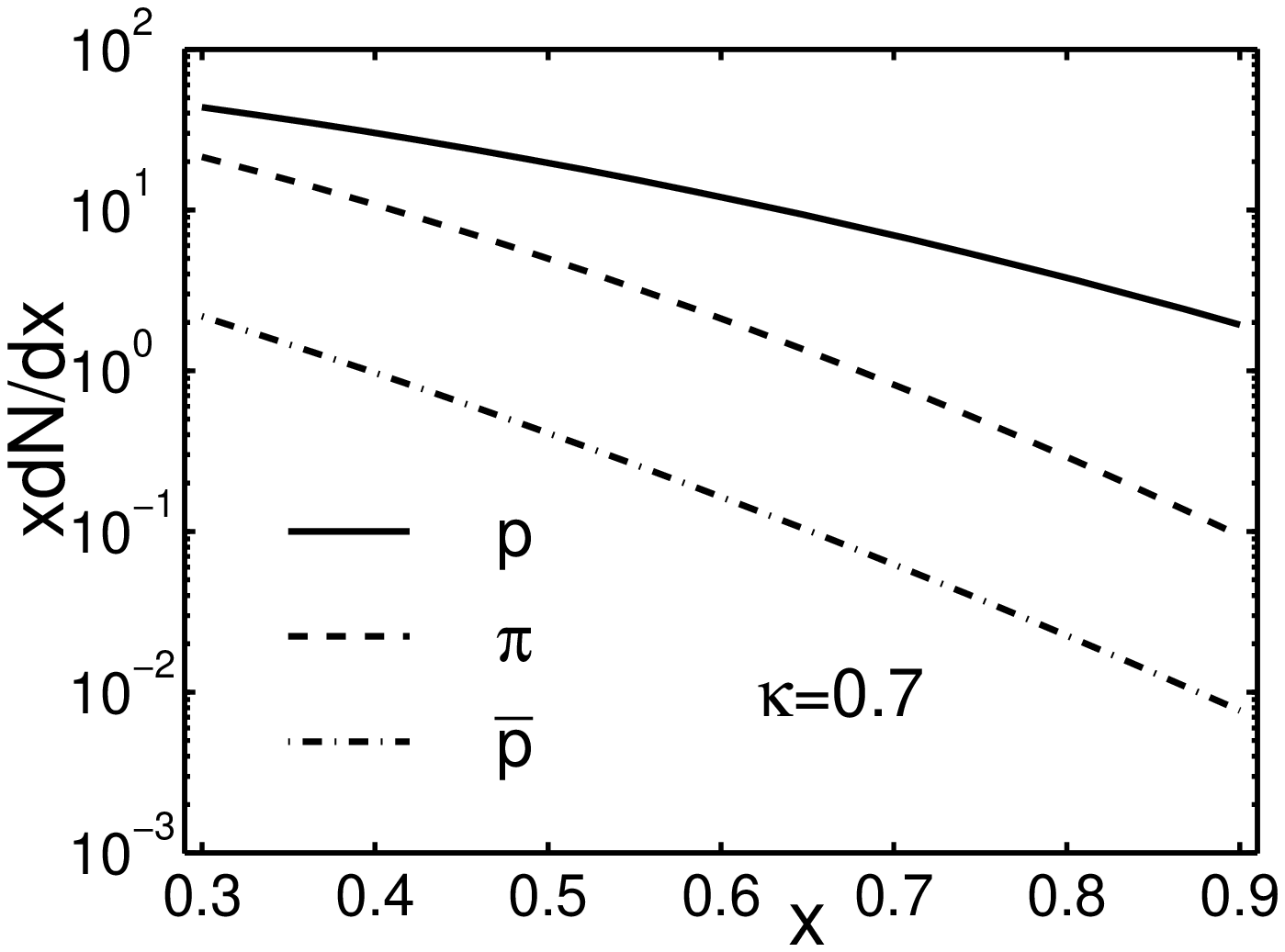}
\caption{
The $x$ \dis s of produced proton, pion and antiproton in Au-Au collisions at $b=3.3$ fm and $\sqrt s=62.4$ GeV for $\kappa=0.7$.}
\end{figure}

\newpage
\begin{figure}[htbp]
\centering
\includegraphics[width=6in]{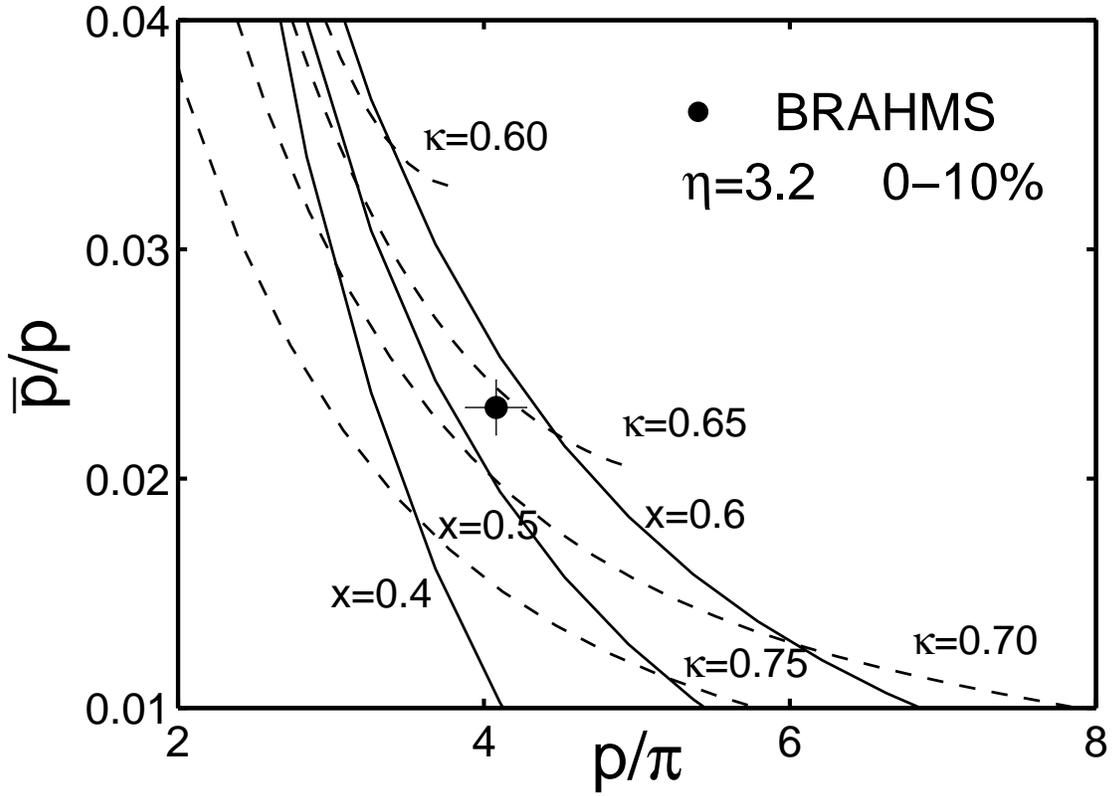}
\caption{
A plot of antiproton/proton ratio versus proton/pion ratio for various fixed values of $x$ (solid lines) and $\kappa$ (dashed lines) for Au-Au collisions at 0-10\% centrality. The theoretical curves are determined by calculating $H_{h'}(x,\kappa)/H_h(x,\kappa)$. The experimental point is from the BRAHMS data on $R_{\bar p/p}$ \cite{ia2} and $R_{p/\pi}$ \cite{ps} at $\sqrt s=62.4$ GeV and $\eta=3.2\pm 0.2$.}
\end{figure}

\newpage
\begin{figure}[htbp]
\centering
\includegraphics[width=6in]{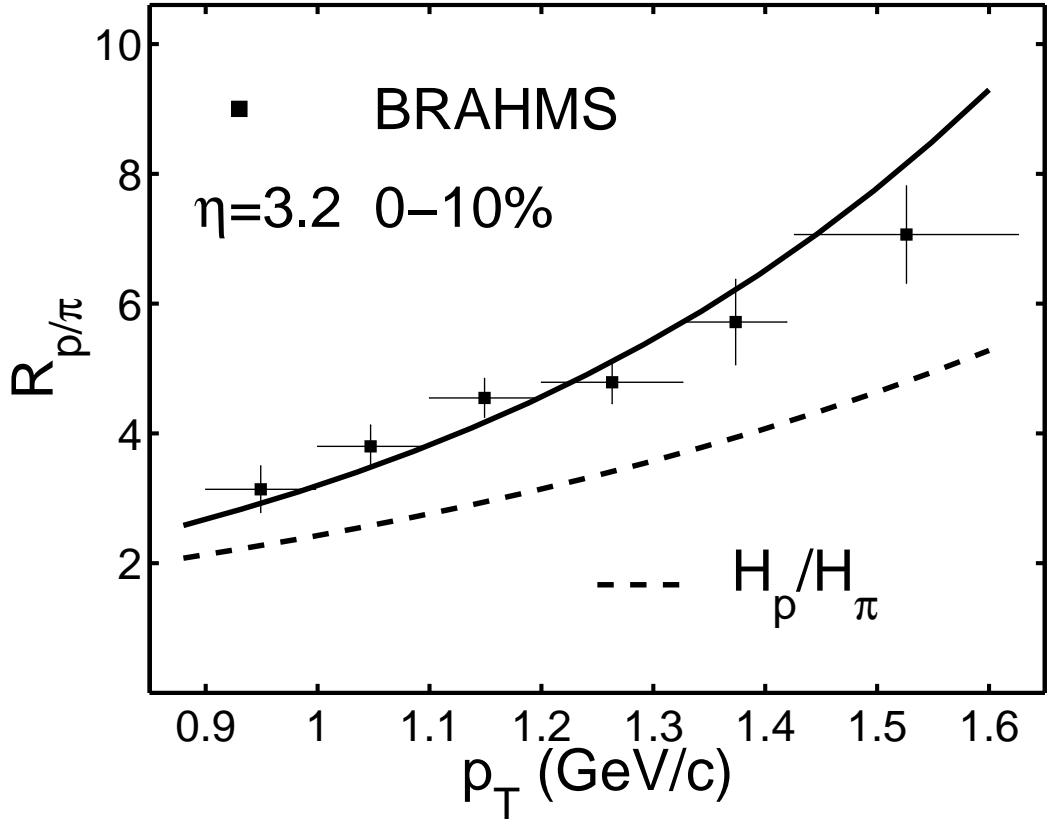}
\caption{
The $\pt$ dependence of proton-to-pion ratio in Au-Au collisions at $\eta=3.2$. The dashed line is obtained from the ratio $H_{p}[x(\pt)]/H_\pi[x(\pt)]$ with $\kappa=0.67$. The solid line includes the factor $V_{p}(\pt)/V_\pi(\pt)$. The data (preliminary) are from \cite{ps}.}
\end{figure}

\newpage
\begin{figure}[htbp]
\centering
\includegraphics[width=6in]{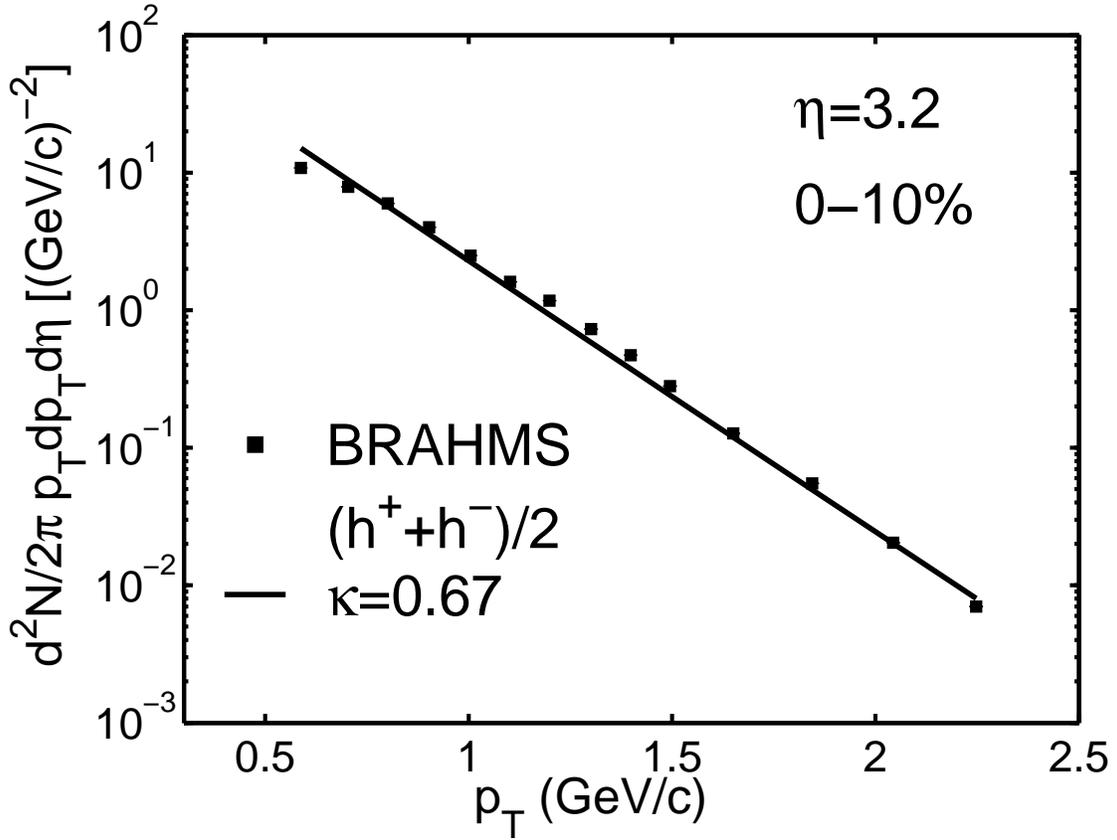}
\caption{
The $\pt$ \dis\ of average charged hadron in Au-Au collisions at $\eta=3.2$. The data are from \cite{ia}. The solid line is obtained by use of Eq.\ (\ref{34}) for $h^{\pm}=[p+\bar p+1.2(\pi^++\pi^-)]/2$ and $\kappa=0.67$. }
\end{figure}

\newpage
\begin{figure}[htbp]
\centering
\includegraphics[width=6in]{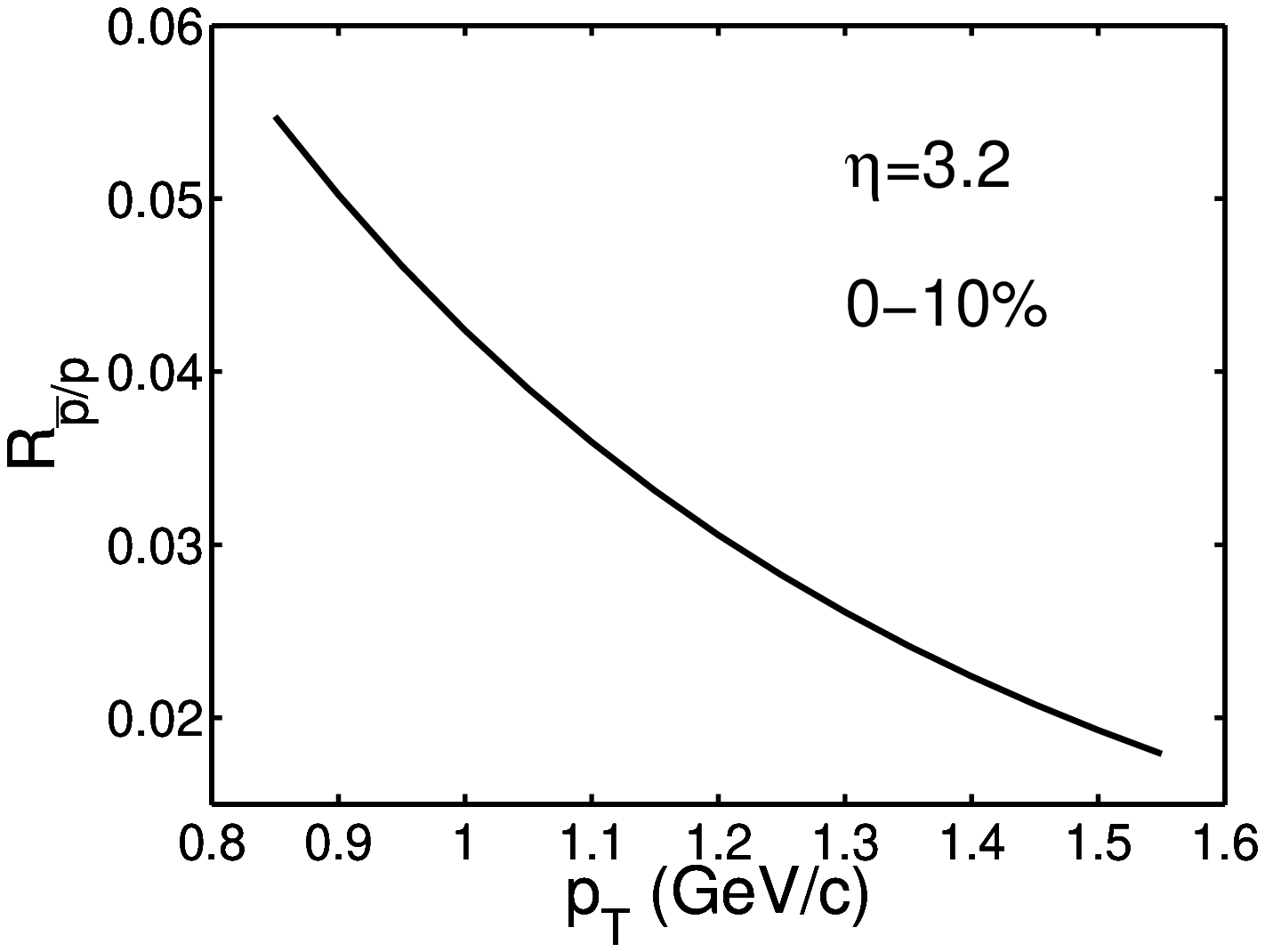}
\caption{
The $\pt$ dependence of antiproton-to-proton ratio for 0-10\% centrality in Au-Au collisions at ${\sqrt s}=62.4$ GeV, $\eta=3.2$  and $\kappa=0.67$.  }
\end{figure}


\newpage
 
\begin{thebibliography}{000}

\bibitem{dh}
K.\ P.\ Das and R.\ C.\ Hwa, Phys.\ Lett.\ {\bf 68B}, 459, (1977).

\bibitem{rch}
R.\ C.\ Hwa, Phys.\ Rev.\ D{\bf 22}, 759 (1980); {\bf 22}, 1593
(1980).

\bibitem{hy3}
R.\ C.\ Hwa and C.\ B.\ Yang, Phys.\ Rev.\ C {\bf 66}, 025205
(2002).

\bibitem{hy}
R.\ C.\ Hwa and C.\ B.\ Yang, Phys.\ Rev.\ C {\bf 73}, 044913
(2006).

\bibitem{hy2}
R.\ C.\ Hwa and C.\ B.\ Yang, Phys.\ Rev.\ C {\bf 76}, 014901 (2007).

\bibitem{phob} 
B.\ B. Back {\it et al.}\ (PHOBOS Collaboration), Phys.\ Rev.\
Lett.\ {\bf 91}, 052303 (2003); Phys.\ Rev.\ Lett.\ {\bf 87}, 102303 (2001).

\bibitem{ia}
I.\ C.\ Arsene {\it et al.}\ (BRAHAMS Collaboration),
nucl-ex/0602018.

\bibitem{ps}
N.\ Katry\'nska and P.\ Staszel (for BRAHAMS Collaboration), poster presentation at Quark Matter 2008, Jaipur, India, arXiv: 0806.1162.

\bibitem{ia2}
I.\ C.\ Arsene (for BRAHAMS Collaboration), Quark Matter 2008, talk presented at Quark Matter 2008, Jaipur, India, arXiv: 0806.0745.

\bibitem{hyc}
H.\ Yang (for BRAHAMS Collaboration), Czech J.\ Phys.\ {\bf 56}, A27 (2006). 

\bibitem{hy4}R.\ C.\ Hwa and C.\ B.\ Yang, Phys.\ Rev.\ C {\bf 65},
034905 (2002).

\bibitem{hy5}R.\ C.\ Hwa and C.\ B.\ Yang, Phys.\ Rev.\ C {\bf 66},
025204 (2002).

\bibitem{hy6}
R.\ C.\ Hwa and C.\ B.\ Yang, Phys.\ Rev.\ C {\bf 70},
024905 (2004).

\bibitem{hy7}
R.\ C.\ Hwa and C.\ B.\ Yang, Phys.\ Rev.\ Lett.\ {\bf 97},
042301 (2006).

\end{thebibliography}
\end{document}